\documentstyle[12pt]{article}

\begin{document}

\begin{flushright}
IMSc/2002/09/34 \\
hep-th/0209129 
\end{flushright} 

\vspace{2mm}

\vspace{2ex}

\begin{center}
{\large \bf Classical Velocity in $\kappa-$deformed Poincare } \\ 

\vspace{2ex}

{\large \bf Algebra and a Maximum Acceleration } \\

\vspace{8ex}

{\large  S. Kalyana Rama}

\vspace{3ex}

Institute of Mathematical Sciences, C. I. T. Campus, 

Taramani, CHENNAI 600 113, India. 

\vspace{1ex}

email: krama@imsc.ernet.in \\ 

\end{center}

\vspace{6ex}

\centerline{ABSTRACT}
\begin{quote} 
We study the commutators of the $\kappa-$deformed Poincare
Algebra ($\kappa$PA) in an arbitrary basis. It is known that the
two recently studied doubly special relativity theories
correspond to different choices of $\kappa$PA bases. We present
another such example. We consider the classical limit of
$\kappa$PA and calculate particle velocity in an arbitrary
basis. It has standard properties and its expression takes a
simple form in terms of the variables in the Snyder basis. We
then study the particle trajectory explicitly for the case of a
constant force. Assuming that the spacetime continuum, velocity,
acceleration, etc. can be defined only at length scales greater
than $x_{min} \ne 0$, we show that the acceleration has a finite
maximum.
\end{quote}

\vspace{2ex}


\newpage

\vspace{4ex}

{\bf 1.}  
Recently, a Lorentz invariant length scale has been incorporated
in the so called doubly special relativity (DSR) theories. It
was originally formulated in the theory, dubbed DSR1, of
\cite{camelia}. A second version, dubbed DSR2, has also been
proposed recently \cite{magu}. Soon after, it has been found in
\cite{glik,luk} that they both can be incorporated within the
framework of $\kappa-$deformed Poincare algebra ($\kappa$PA)
\cite{kappa}, and correspond to different choices of $\kappa$PA
bases.  Lorentz transformation and addition properties of the
energy momentum vectors have also been found \cite{judes}, and
incorporated within the $\kappa$PA framework \cite{glik2}.

As shown in \cite{glik}, the commutators of the $\kappa$PA can
be expressed succinctly in the Snyder basis \cite{snyder}. The
energy, momentum, and mass in this basis can be related to those
in another basis by three arbitrary functions in general.
Lorentz transformation and addition properties of the energy
momentum vectors in these bases are then related. For details,
see \cite{glik,luk,judes,glik2}.

Many DSR theories can be constructed within the $\kappa$PA
framework, characterised by three functions. In the present
paper, we exhibit one such example by making a particular choice
of these functions, chosen to reproduce the commutators in the
position momentum sector studied in \cite{maggiore,k,maggiore2}
in the context of generalised uncertainty principle.

There are many possible ways to define the velocity in these
theories \cite{lukvel}. Following \cite{lukvel}, we study the
classical limit of $\kappa$PA, where the commutators are
replaced by Poisson brackets. Velocity, acceleration, etc.  can
then be defined naturally. We compute the velocity in any basis,
characterised by the three arbitrary functions. Its expression
takes a simple form in terms of the variables in the Snyder
basis which also shows that the maximum speed, as also the speed
of a massless particle, is given by the speed of light in
vacuum. With a natural assumption regarding the Lorentz
transformation of energy and momentum, the velocities can be
shown to obey the standard relativistic addition law.

We also study the motion of a particle under the action of a
potential obtaining, in particular, the particle trajectory
explicitly for the case of a constant force. Assuming that the
spacetime continuum, velocity, acceleration, etc. can be defined
only at length scales greater than $x_{min} \ne 0$, we show that
the acceleration has a finite maximum. Our arguments can be
extended straightforwardly, leading to the same conclusion, in
any theory where the above assumption is valid. An upper bound
on acceleration has also been obtained previously in other
theories \cite{maxacc,giddings} and studied further in
\cite{lamb}.

The plan of the paper is as follows. In section {\bf 2}, we
present the commutators of the $\kappa$PA in a general basis,
the choices of functions leading to DSR1 and DSR2, and another
example of DSR. In section {\bf 3}, we study the classical limit
and obtain the velocity. In section {\bf 4}, we present the
particle motion under the action of a constant force and obtain
an upper bound on the acceleration. We conclude in section {\bf
5} with a brief summary and a mention of a few issues for
further studies.

\vspace{4ex}

{\bf 2.}  
Let $X^\mu$ and $K_\mu$, $\mu = (0, i)$, $i = 1, 2, \ldots, d$,
denote the position and momentum operators, and $M_{\mu \nu}$,
explicitly $M_{i 0} \equiv N_i$ and $M_{i j}$, denote the boost
and angular momentum generators in $(d + 1)-$dimensional
spacetime. Also, let $\eta_{\mu \nu} = diag (- 1, 1, 1,
\ldots)$, and $\hbar = c = 1$. Then, in the Snyder basis
\cite{snyder} for the $\kappa-$deformed Poincare Algebra
($\kappa$PA) the commutators of the above operators can be
written succinctly \cite{glik} as ${[} K_\mu, K_\nu {]} = 0$ and 
\begin{eqnarray}
{[} M_{\mu \nu}, M_{\rho \sigma} {]} & = & 
i \; (\eta_{\mu \rho} M_{\nu \sigma} 
- \eta_{\nu \rho} M_{\mu \sigma}
- \eta_{\mu \sigma} M_{\nu \rho}
+ \eta_{\nu \sigma} M_{\mu \rho}) \nonumber \\
{[} M_{\mu \nu}, V_\sigma {]} & = & 
i \; (\eta_{\mu \sigma} V_\nu 
- \eta_{\nu \sigma} V_\mu) \; , \; \; \; 
V_\sigma = X_\sigma, \; K_\sigma \nonumber \\
{[} X_\mu, K_\nu {]} & = & i \; 
(\eta_{\mu \nu} + \alpha K_\mu K_\nu) \; , \; \; \; 
{[} X_\mu, X_\nu {]} = i \; \alpha M_{\mu\nu} \label{1} 
\end{eqnarray}
where $\alpha = \lambda^2 = \kappa^{- 2}$ is the deformation
parameter, with dimension $(length)^2$. The Casimir ${\cal C}_s$
in the Snyder basis, which commutes with $N_i$, is given by 
\begin{equation}\label{cs} 
{\cal C}_s = K_0^2 - K^2 
\; \; \; {\rm where} \; \; \; 
K^2 = \sum_{i = 1}^d K_i^2 \; .
\end{equation}
When $\alpha \ne 0$, there is no clear physical criterion to 
identify the physical momentum operator, $P_\mu$. Various definitions 
of $P_\mu (K_0, K_i)$ have been employed, which amount to choosing 
different bases for the $\kappa$PA \cite{glik,luk}. Thus, let 
\begin{equation}\label{gh}
P_0 = h(K_0, K) 
\; , \; \; \; 
P_i = g(K_0, K) K_i 
\end{equation}
where $h \to K_0$ and $g \to 1$ in the limit $\alpha \to 0$, and
the above relations can be inverted to obtain $K_0(P_0, P_i)$
and $K_i(P_0, P_i)$. It is then straightforward to obtain the
new commutators. Thus, the commutators involving $P_\mu$ become
${[} M_{i j}, P_k {]} = i (\delta_{i k} P_j - \delta_{j k}
P_i)$, ${[} M_{i j}, P_0 {]} = 0$, and
\begin{eqnarray} 
{[} N_i, P_j {]} & = & i \; \left( 
g \delta_{i j} + \left(
\frac{g_{_{K_0}}}{K_0} + \frac{g_{_K}}{K} 
\right) \; K_i K_j \right) K_0 \nonumber \\
{[} X_i, P_j {]} & = & i \; \left( 
g \delta_{i j} + \left( \frac{g_{_K}}{K} 
+ \alpha (g + K_0 g_{_{K_0}} + K g_{_K})
\right) \; K_i K_j \right) \nonumber \\
{[} X_0, P_i {]} & = & i \; \left( 
- \frac{g_{_{K_0}}}{K_0}
+ \alpha (g + K_0 g_{_{K_0}} + K g_{_K})
\right) \; K_0 K_i \nonumber \\
{[} N_i, P_0 {]} & = & i \; \left( 
\frac{h_{_{K_0}}}{K_0} + \frac{h_{_K}}{K} 
\right) \; K_0 K_i \nonumber \\
{[} X_i, P_0 {]} & = & i \; \left( \frac{h_{_K}}{K} 
+ \alpha (K_0 h_{_{K_0}} + K h_{_K}) 
\right) \; K_i \nonumber \\
{[} X_0, P_0 {]} & = & i \; \left( 
- \frac{h_{_{K_0}}}{K_0}
+ \alpha (K_0 h_{_{K_0}} + K h_{_K}) 
\right) \; K_0  \label{pnx}
\end{eqnarray} 
where, here and in the following, a subscript denotes partial
differentiation with respect to it. The commutators in (\ref{1})
are obtained by setting $g = 1$ and $h = K_0$. It is also easy
to verify that the Casimir ${\cal C}(P_0, P_i)$ in the new
basis, which commutes with $N_i$, is given by 
\begin{equation}\label{c}
\alpha {\cal C} = {\cal F}(\alpha (K_0^2 - K^2)) 
\; \; \; {\rm with} \; \; \; {\cal F}(0) = 0 
\; \; \; {\rm and} \; \; \; {\cal F}'(0) = 1 
\end{equation}
where ${\cal F}'$ is the derivative of ${\cal F}$ with respect
to its argument. The restriction on ${\cal F}$ and ${\cal F}'$
ensures that the correct Casimir is obtained in the limit
$\alpha \to 0$. The function ${\cal F}$ is otherwise
arbitrary. In equations (\ref{pnx}) and (\ref{c}), $K_0$ and
$K_i$ are to be expressed in terms of $P_0$ and $P_i$. The mass
shell condition ${\cal C} = m^2$ gives the Hamiltonian $H =
P_0$, equivalently energy $E$, as a function of $P$, $m$, and
$\alpha$, and also that $\alpha m^2 = {\cal F}(\alpha m_s^2)$
where ${\cal C}_s = m_s^2$ in the Snyder basis. Thus, the
function ${\cal F}$ defines $m$ in terms of $m_s$.

Under a Lorentz transformation, let $K_\mu$ transform in the
standard way {\em i.e.} $K_\mu \to K'_\mu = \Lambda_\mu \; ^\nu
\; K_\nu$. Then, the transformation $P_\mu \to P'_\mu$ can be
taken to be given by 
\begin{equation}
P_0' = h(K'_0, K'_i) \; , \; \; \; 
P_i' = g(K'_0, K'_i) \; K'_i \label{k'p'} \; . 
\end{equation}
Similarly, let the addition law for $K_\mu$ be the standard one,
{\em i.e.} $K^{total}_\mu \equiv K'_\mu = \sum_a K^{(a)}_\mu$.
Then, the addition law for $P_\mu$ can be taken to be given by
$P^{total}_\mu \equiv P'_\mu$, with $P'_\mu$ given by 
(\ref{k'p'}) \cite{glik,luk,judes,lukvel}. 

In the above formulation, with $\alpha = \lambda^2$, the DSR
theories \cite{camelia,magu} correspond to different choices of
the functions $g$, $h$, and ${\cal F}$. Thus, for DSR1, $P_\mu$,
$K_\mu$, the Casimir ${\cal C}$, and the function ${\cal F}$
in(\ref{c}) are given by 
\begin{eqnarray}
e^{\lambda P_0} & = & \lambda K_0 + 
\sqrt{1 + \lambda^2 (K_0^2 - K^2)} \; , \; \; \; 
P_i = K_i e^{- \lambda P_0} \nonumber \\
\lambda K_0 & = & Sinh \lambda P_0 
+ \frac{\lambda^2 P^2}{2} e^{\lambda P_0} \; , \; \; \; 
\; \; \;   \; \; \;   \; 
K_i = P_i e^{\lambda P_0} \nonumber \\
\lambda^2 {\cal C} & = & 2 (\sqrt{1 
+ \lambda^2 (K_0^2 - K^2)} - 1) 
= (2 Sinh \frac{\lambda P_0}{2})^2  
- \lambda^2 P^2 e^{\lambda P_0}  \label{dsr1}
\end{eqnarray} 
where $P^2 = \sum P_i^2$. For DSR2, they are given by
$\frac{P_i}{P_0} = \frac{K_i}{K_0}$ and    
\begin{eqnarray}
P_0 & = & \frac{K_0}{\lambda K_0 + 
\sqrt{1 + \lambda^2 (K_0^2 - K^2)}} \; , \; \; \; 
K_0 = \frac{P_0}{\sqrt{1 - 2 \lambda P_0 
+ \lambda^2 P^2}} \nonumber \\
\lambda^2 {\cal C} & = & \frac{\lambda^2 (K_0^2 - K^2)}
{1 + \lambda^2 (K_0^2 - K^2)} = \frac{\lambda^2 
(P_0^2 - P^2)}{(1 - \lambda P_0)^2} \; . \label{dsr2}
\end{eqnarray} 
The Lorentz transformation and the addition laws for $P_\mu$
follow from (\ref{k'p'}). See \cite{camelia,magu,glik,luk,judes}
for more details.

Consider the ${[}X_i, K_j{]}$ commutators, of the form
\begin{equation}\label{ab}
{[} X_i, K_j {]} = i \; (A(K) \delta_{i j} + B(K) K_i K_j) \; , 
\end{equation} 
which appear in the context of generalised uncertainty
principle, studied in \cite{maggiore,k,maggiore2,kempf}. Let
$P_i = g(K) K_i$. One then gets 
\begin{equation}\label{abtilde}
{[} X_i, K_j {]} = i \; (\tilde{A} 
\delta_{i j}  + \tilde{B} K_i K_j) 
\end{equation}
where $\tilde{A}$ and $\tilde{B}$ are given by 
\begin{equation}\label{btilde}
\tilde{A} = g A \; , \; \; \; 
P^2 \tilde{B} = A K g_{_K}
+ B K^2 (g + K g_{_K}) \; . 
\end{equation}
For $A = 1$ and $B = 0$, for example, one has $\tilde{A} = g$
and $\tilde{B} = \frac{2 g g_{_{P^2}}}{g - 2 P^2 g_{_{P^2}}}$,
which is the case studied in \cite{kempf}. Let $A = 1$ and $B =
\alpha$, which corresponds to the ${[}X_i, K_j{]}$ commutators 
in (\ref{1}) in the Snyder basis. Furthermore, let $g$ be 
chosen such that $\tilde{B} = 0$. It then follows that 
\begin{equation}\label{g}
\tilde{A} = g \; , \; \; \; 
g^2 = \frac{P^2}{K^2} = \frac{a^2}{1 + \alpha K^2} 
= a^2 - \alpha P^2 
\end{equation}
where $a$ is an integration constant. For $a^2 = 1 - \alpha m^2$
and $\alpha = \pm \lambda^2$, this is the case studied in
\cite{maggiore,k,maggiore2}.

A DSR theory can be constructed within the frameowrk of
$\kappa$PA by choosing functions $g$, $h$, and ${\cal F}$ in
equations (\ref{gh}) and (\ref{c}). This was illustrated above
for DSR1 and DSR2. We now present another example. Let $h = g
K_0$ and the function $g$ be given by equation (\ref{g}) with $a
= 1$ and $\alpha = \lambda^2$. Then, one obtains
$\frac{P_i}{P_0} = \frac{K_i}{K_0}$ and 
\begin{equation}\label{dsr3}
P_0 = \frac{K_0}{\sqrt{1 + \lambda^2 K^2}} 
\; , \; \; \; 
K_0 = \frac{P_0}{\sqrt{1 - \lambda^2 P^2}} \; . 
\end{equation}
The Casimir depends on the choice of ${\cal F}$. For example, 
\begin{equation}\label{f3} 
\lambda^2 {\cal C} = \lambda^2 (K_0^2 - K^2) = 
\frac{\lambda^2 (P_0^2 - P^2)}{1 - \lambda^2 P^2} \; . 
\end{equation}
The Lorentz transformation and the addition laws for $P_\mu$
follow from (\ref{k'p'}).

Note that, when $\lambda = 0$, the Casimir ${\cal C}$, and thus
the mass shell condition ${\cal C} = m^2$, is symmetric under
$P_\mu \longleftrightarrow - P_\mu$. However, when $\lambda \ne
0$, this symmetry is absent for DSR1 and DSR2 (see equations
(\ref{dsr1}) and (\ref{dsr2})) but is present for the example
presented above (see equations (\ref{f3})).

\vspace{4ex}

{\bf 3.}  
The mass shell condition ${\cal C} = m^2$ gives the Hamiltonian
$H(P) = P_0$ as a function of $P$, $m$, and $\alpha$. The
corresponding eigenvalues then give the energy $E(p)$. Now,
there are many possible ways to define the velocity. For
example, $v_i$ can be defined as $\frac{\partial E}{\partial
p_i}$, or as the eigenvalues of the operator $i {[}H, X_i{]}$,
or, in $\kappa$PA, as the left$-$ or right$-$ covariant
velocities \cite{lukvel}. See \cite{tamaki} also. 

In the classical limit the commutators are replaced by the
Poisson brackets. Given a Hamiltonian function ${\cal H}(Y_a)$,
where $Y_a = (x_\mu, p_\mu)$, the evolution of an observable $O$
with respect to an evolution parameter $s$ is given by
\begin{equation}\label{os} 
\frac{d O}{d s} = {[}O, Y_a{]}_{PB} \; 
\frac{\partial {\cal H}}{\partial Y_a} \; . 
\end{equation} 
Then, velocity $v_i$ and acceleration $a_i$ can be defined
naturally to be given by 
\begin{equation}\label{v} 
v_i = \frac{d x_i/d s}{d x_0/d s} \; , \; \; \; 
a_i = \frac{d v_i/d s}{d x_0/d s} \; . 
\end{equation} 
The interpretation of $s$ depends on the choice of 
${\cal H}$. For example, if 
\begin{equation}\label{calh}
{\cal H} = p_0 - E(p) \; , 
\end{equation}
then, in the limit $\alpha \to 0$, the parameter $s$ can be
identified with time \cite{lukvel}. 

It has been shown in \cite{lukvel}, for the case of DSR1, that
the velocity given by (\ref{v}) is the same as the
right$-$covariant velocity in the bicross product basis of the
$\kappa$PA. Also, the maximum speed $v_{max} = 1$ and the
velocities satisfy the standard relativistic addition law.

In the following, we take the velocity $v_i$ and the Hamiltonian
function ${\cal H}$ to be given by (\ref{v}) and (\ref{calh}),
and compute $v_i$ for the general case specified by the
arbitrary functions $g(k_0, k)$, $h(k_0, k)$, and ${\cal
F}(\alpha(k_0^2 - k^2))$, defined in equations (\ref{gh}) and
(\ref{c}).

Using the Poisson brackets obtained from the commutators
(\ref{pnx}), and after some algebra, the velocity $v_i$ given by
(\ref{v}) can be written explicitly as
\begin{equation}\label{vexp}
v_i = \frac{k_i}{k_0} \; \left( 
\frac{{\cal A}}{{\cal A} - {\cal B}} \right) 
\end{equation}
where 
\begin{eqnarray}
{\cal A} & = & \frac{E_p}{k} \; \left( g + k g_{_k} 
+ \alpha k^2 (g + k_0 g_{_{k_0}} + k g_{_k}) \right) 
- \left( \frac{h_{_k}}{k} + \alpha 
(k_0 h_{_{k_0}} + k h_{_k}) \right) \nonumber \\
{\cal B} & = & 
\frac{h_{_{k_0}}}{k_0} + \frac{h_{_k}}{k} 
- \frac{E_p}{k} \; \left( g + k g_{_k} 
+ \frac{k^2 g_{_{k_0}}}{k_0} \right) 
\; , \label{abcal} 
\end{eqnarray} 
and $k_0$, $k$, ${\cal A}$, ${\cal B}$, etc. are all functions
of $p_0$ and $p$. Now, given the function ${\cal F}$, the
Casimir given by (\ref{c}) can be written as 
\begin{equation}\label{FG}
F^2(p_0, p) - G^2(p_0, p) \; p^2 = 
{\cal F}(\alpha (k_0^2 - k^2)) \; ,
\end{equation}
the precise form of $F$ and $G$ depending on the choice of
functions $g$, $h$, and ${\cal F}$. Using $p_0 = h(k_0, k)$ and
performing the operation $\left( \frac{1}{k_0} \;
\frac{\partial}{\partial k_0} + \frac{1}{k} \;
\frac{\partial}{\partial k} \right)$ on both sides of equation
(\ref{FG}) gives 
\begin{equation}\label{FG1}
(F^2 - G^2 p^2)_h \; \left( 
\frac{h_{_{k_0}}}{k_0} + \frac{h_{_k}}{k} \right) 
+ (F^2 - G^2 p^2)_p \; \left( 
\frac{p_{_{k_0}}}{k_0} + \frac{p_{_k}}{k} \right) = 0 \; . 
\end{equation}
The energy $E(p) = p_0 = h$ is obtained from the mass shell
condition ${\cal C} = m^2$. Thus, $F^2(E, p) - G^2(E, p) \; p^2
= \alpha m^2$, differentiating which gives 
\begin{equation}\label{ep}
(F^2 - G^2 p^2)_h \; E_p + (F^2 - G^2 p^2)_p = 0 \; . 
\end{equation}
Furthermore, using $p_i = g(k_0, k) k_i$ one obtains 
$p = k g(k_0, k)$ and, hence, 
\begin{equation}\label{pk}
\frac{p_{_{k_0}}}{k_0} + \frac{p_{_k}}{k} = 
\frac{1}{k} \; \left( g + k g_{_k} 
+ \frac{k^2 g_{_{k_0}}}{k_0} \right) \; . 
\end{equation} 
Equation (\ref{FG1}) can now be written, using equations
(\ref{ep}) and (\ref{pk}), as 
\begin{equation}\label{b=0}
(F^2 - G^2 p^2)_h \; {\cal B} = 0
\end{equation} 
where ${\cal B}$ is given in (\ref{abcal}). Since $F$ and $G$
are arbitrary functions, it follows that ${\cal B} = 0$. 
Then, equation (\ref{vexp}) becomes simply 
\begin{equation}\label{vfinal}
v_i = \frac{k_i(p_0, p)}{k_0(p_0, p)} 
\end{equation}
which is valid for the general case specified by the arbitrary
functions $g(k_0, k)$, $h(k_0, k)$, and ${\cal F}(\alpha(k_0^2 -
k^2))$. Note that $k_0$ and $k_i$ are the momentum varaibles in
the Snyder basis, whereas $p_0$ and $p_i$ are those in a
particular basis under consideration. Thus, using equations
(\ref{dsr1}), (\ref{dsr2}), and (\ref{dsr3}), $v_i$ for the case
of DSR1 \cite{lukvel,tamaki}, DSR2 \cite{magu}, and the present
example are given by 
\begin{equation}\label{veldsr1}
v_i \vert_{_{DSR1}} = \frac{2 \lambda p_i}{1 - e^{- 2 \lambda p_0} 
+ \lambda^2 p^2} \; , \; \; \;  
v_i \vert_{_{DSR2}} = 
v_i \vert_{_{Example}} = \frac{p_i}{p_0} \; . 
\end{equation}
Equation (\ref{vfinal}) implies that the maximum speed $v_{max}
= 1$, as is the speed of a massless particle.\footnote{ However,
if $v_i$ is given, for example, by $\frac{\partial E}{\partial
p_i}$ or the eigenvalues of the operator $i {[}H, X_i{]}$ then
$v_{max} \ne 1$ in general. In fact, the functions $g$ and $h$
can be chosen such that $v_{max} > 1$.} Also, using equation
(\ref{k'p'}) and following \cite{lukvel}, it is straightforward
to prove that the velocities defined above satisfy the standard
relativistic addition law for any choice of functions $g$, $h$,
and ${\cal F}$.

\vspace{4ex}

{\bf 4.}  
We now study the motion of a particle under the action of a
potential in the classical limit of the $\kappa$PA, with the
Hamiltonian function ${\cal H}(x, p)$ given by 
\begin{equation}\label{hv}
{\cal H} = p_0 - E(p) - V(x) 
\end{equation}
where the potential $V(x)$ is a function of $x^i$ only. For this
purpose, it is convenient to work in the bicross product basis.
With $\alpha = \lambda^2$, the relevent Poisson brackets in this
basis are given by \cite{lukvel} 
\begin{eqnarray}
{[}p_i, x_j{]}_{PB} = \delta_{i j} \; , \; \; 
& {[}p_0, x_i{]}_{PB} = 0 \; , \; \; & 
{[}x_0, x_i{]}_{PB} = \lambda x_i \nonumber \\
{[}p_i, x_0{]}_{PB} = \lambda p_i \; , \; \; 
& {[}p_0, x_0{]}_{PB} = - 1 \; , \; \; & 
{[}p_\mu, p_\nu{]}_{PB} = 0 \; . \label{poisson}
\end{eqnarray}
The evolution of an observable $O(s)$ is given by equation
(\ref{os}) where $s$ is an evolution parameter which, in the
limit $\lambda \to 0$, can be identified with time. Equation
(\ref{os}) determines, in particular, that the evolution of
$p_i(s)$, $x_i(s)$, and $x_0(s)$ is given by the differential
equations 
\begin{equation}\label{pxix0}
\frac{d p_i}{d s} = f_i(x) \; , \; \; \; 
\frac{d x_i}{d s} = u_i(p) \; , \; \; \; 
\frac{d x_0}{d s} = D(x, p)
\end{equation}
where we have defined
\[
f_i = - \frac{\partial V(x)}{\partial x_i} 
\; , \; \; \; 
u_i = \frac{\partial E(p)}{\partial p_i} 
\; , \; \; \; 
D = 1 + \lambda (p_j u_j + x_j f_j) \; . 
\]  
Note that $f_i$ is the $i^{th}$ component of the force due to
the potential $V(x)$ and, in the limit $\lambda \to 0$, $u_i$ is
the $i^{th}$ component of the velocity. Also, it follows that 
\begin{equation}\label{e+v}
\frac{d}{d s} \; \left(E(p) + V(x)\right) = 0 
\; \; \; \longleftrightarrow \; \; \; 
E(p) + V(x) = constant \; . 
\end{equation}
The velocity $v_i$ and acceleration $a_i$, defined in (\ref{v}),
are now given by 
\begin{eqnarray}
v_i & = & \frac{u_i}{D} \label{vel} \\
a_i & = & \frac{f_j}{D^2} \; 
\left( \frac{\partial u_i}{\partial p_j} \right) 
- \frac{\lambda u_i f_j}{D^3} \; \left( 2 u_j 
+ p_k \frac{\partial u_k}{\partial p_j} \right)  
- \; \frac{\lambda u_i u_j x_k}{D^3} \; \left(
\frac{\partial f_k}{\partial x_j} \right) \; . 
\label{acc} 
\end{eqnarray}

Solutions to equations (\ref{pxix0}) will describe the motion of
a particle under the action of a potential $V(x)$ in the
classical limit of the $\kappa$PA. Note that, generically, there
are two arbitrary functions, namely $E(p)$ and $V(x)$, and
solutions to equations (\ref{pxix0}) are difficult to obtain in
the general case.

Let $V(x) = - f x_1$ with $f = constant$, which corresponds to a
constant force, chosen to be along $x_1$ with no loss of
generality. Let $x_0 = x_i = p_i = 0$ at $s = 0$. Then, it
follows that $x_i = \delta_{i 1} x$ and $p_i = \delta_{i 1} p$.
Equations (\ref{pxix0}) can be solved easily, with the resulting
solution given simply by 
\begin{equation}\label{solution}
p = s f \; , \; \; \; 
x = \frac{E(p) - E(0)}{f} \; , \; \; \; 
x_0 = s \; (1 + \lambda x f) \; , 
\end{equation}
which is valid for any function $E(p)$.  Note that equation
(\ref{e+v}) is also satisfied.

To obtain a complete solution, $E(p)$ must be specified. As an
example, consider the case of DSR1 with the Casimir ${\cal C}$,
equivalently the function ${\cal F}$, given in (\ref{dsr1}).
The energy $E(p)$, obtained by setting ${\cal C} = m^2$ with
$p_0 = E(p)$, is given by 
\[ 
e^{- \lambda E} = b - \sqrt{b^2 - (1 - \lambda^2 p^2)} 
\] 
where $b = 1 + \frac{\lambda^2 m^2}{2}$. Equations
(\ref{solution}) now determine $p(s)$, $x(s)$, and $x_0(s)$
completely.

As follows from equations (\ref{1}) and (\ref{poisson}), the
spacetime is noncommutative at length scales smaller than ${\cal
O}(1) \; \lambda$. Therefore, spacetime continuum, velocity,
acceleration etc. can be defined only at length scales greater
than $x_{min} = {\cal O}(1) \; \lambda$. Hence, solutions
(\ref{solution}) are also applicable only for $x \ge x_{min}$.

Consider a particle with mass $m$ such that $\lambda m \ll
1$. Its motion under a constant force is given by equations
(\ref{solution}), taken to be valid for $x \ge x_{min}$ only;
equivalently, in the limit $\lambda m \ll 1$, for $s \ge s_{min}
= \left( \frac{2 m}{f} \; x_{min} \right)^{\frac{1}{2}}$ only.
The acceleration at $s_{min}$ is given by 
\begin{equation}\label{acc1}
a_{cont}(s_{min}) = \frac{m^2 f}
{(m^2 + 2 m f x_{min})^{\frac{3}{2}}} \; (1 + \cdots) \; , 
\end{equation} 
where $\cdots$ represents terms which are negligible in the
limit $\lambda m \ll 1$, and $a_{cont}(s)$, $s \ge s_{min}$, is
the acceleration of the particle under a force $f$, which can be
defined and measured in the conventional way using the concept
of spacetime continuum.

Consider $a_{cont}$ as a function of the applied force $f$. If
$\lambda = 0$ then $x_{min} = 0$ and the maximum value of the
acceleration $a_{cont}$ is infinite, achieved when $f \to
\infty$. If $\lambda \ne 0$ then $a_{cont}$ has a finite
maximum, $a_{max}$, achieved when $f = f_* = \frac{m}{x_{min}}$.
Upto numerical coefficients of ${\cal O}(1)$, $a_{max}$ is given
by 
\begin{equation}\label{maxacc}
a_{cont} \le a_{max} \simeq \frac{1}{x_{min}} 
\simeq \frac{1}{\lambda} \; , 
\end{equation} 
where we have used $x_{min} = {\cal O}(1) \; \lambda$. The above
expression remains valid also when $\lambda m$ is not
negligible, but with different numerical coefficients.

Note that for $f = f_*$, $p \simeq E \simeq {\cal O}(1) \; m$ at
$s = s_{min}$. This suggests that when $f > f_*$, the excess
force is likely to create more particle$-$antiparticle pairs,
and not increase the particle acceleration beyond $a_{max}$
({\em cf.} footnote 2 below).

Consider a composite of $N$ particles, of average mass $m$ with
$\lambda m \ll 1$. Then $m_{total} = N m$. For such a composite,
it has been proposed that $\lambda$ must be replaced by
$\lambda_{eff} = \frac{\lambda}{N}$ \cite{magu2}. We assume this
to be the case. Since one must still have $x > x_{min} = {\cal
O}(1) \lambda$ in (\ref{solution}), and since $\lambda_{eff}
m_{total} = \lambda m \ll 1$, it follows for a composite object
also that the acceleration $a_{cont}$ must obey the bound given
in (\ref{maxacc}).

Our arguments can be extended straightforwardly to show that an
upper bound on the acceleration exists, and is of the form given
in (\ref{maxacc}), in any theory where there is a non zero
length scale only above which can one define the spacetime
continuum, velocity, acceleration, etc.
 
Note that an upper bound on acceleration has also been obtained
previously in other theories \cite{maxacc} and studied further
in \cite{lamb}. It also follows, through Unruh effect, in
theories with a limiting temperature such as string theory with
Hagedorn temperature \footnote{In string theory, excess energy
pumped in to increase the temperature goes, instead, into
increasing the length of one `long' string.} \cite{giddings}. It
will be interesting to study whether conversely an upper bound
on acceleration, such as that in the present work, implies a
limiting temperature.

\vspace{4ex}

{\bf 5.}  
We now summarise briefly the present work and mention a few
issues for further studies. We consider arbitrary bases for
the $\kappa$PA, characterised by three functions which redefine
energy, momentum, and mass. DSR1 and DSR2 theories are obtained
by particular choices of these functions, and another example of
DSR is presented.

We study the Hamiltonian evolution in the classical limit,
defining velocity $v_i$, acceleration $a_i$, etc. following
\cite{lukvel}. We find that $v_i$ thus defined is given simply
by equation (\ref{vfinal}). Also, with a natural assumption
regarding the Lorentz transformation of momentum and energy
(\ref{k'p'}), the velocities obey the standard relativistic
addition law.

We also study the motion of a particle under the action of a
potential obtaining, in particular, the particle trajectory
explicitly for the case of a constant force. Assuming that the
spacetime continuum, velocity, acceleration, etc. can be defined
only at length scales greater than $x_{min} \ne 0$, we then show
that the acceleration has a finite maximum given in 
(\ref{maxacc}).

We mention a few issues for further studies. For $\alpha =
\lambda^2$, the theory can be viewed as that with a de Sitter
momentum space \cite{glik2}. It may be of interest to study the
case $\alpha = - \lambda^2$ which is likely to be a theory with
anti de Sitter momentum space (see \cite{maggiore2}).

An important issue is to understand the physical significance of
different bases, equivalently whether any particular basis is
preferred physically. It is also equally important to understand
how to describe a composite of a collection of particles within
the framework of $\kappa$PA. See \cite{magu2} for a recent
proposal for such a description.

\vspace{3ex}

{\bf Acknowledgement:} 
We thank G. Date for a discussion and G. Lambiase 
for bringing \cite{lamb} to our attention. 


\end{document}